\documentclass[12pt, a4paper]{article}
\usepackage{a4wide, fancyhdr, lastpage, graphicx, wrapfig, tikz, amsmath, amssymb, amsthm, xfrac, mathtools, calc, bbm, enumerate, verbatim, yhmath, lscape, accents, afterpage, pdflscape, array, tabularx, stmaryrd}
\usepackage{hyperref}

\usepackage{caption, subcaption}
\usepackage{diagbox}
\usepackage{float}
\usepackage{listings}

\usepackage{needspace, dsfont}

\usepackage{times}

\title{Progress in Artificial Intelligence and its Determinants\thanks{Acknowledgments: Michael T. Cusick from Bureau of Economic Analysis, Patrycja Milewska from Bureau of Labor Statistics, Robyn Rosenberg from Harvard Library. S.V. was supported by the Center of Mathematical Sciences and Applications at Harvard University.}}

\author{Michael R. Douglas\thanks{Contact address: \texttt{mdouglas}\texttt{@cmsa.fas.harvard.edu}.} \ and Sergiy Verstyuk\thanks{Contact address: \texttt{verstyuk}\texttt{@cmsa.fas.harvard.edu}.} }

\date{17 January 2025}

\begin{document}

\maketitle 
\thispagestyle{empty}

\begin{abstract}
We study long-run progress in artificial intelligence in a quantitative way.
Many measures, including traditional ones such as patents and publications, machine learning benchmarks, and a new Aggregate State of the Art in ML (or ASOTA) Index we have constructed from these, show exponential growth at roughly constant rates over long periods.
Production of patents and publications doubles every ten years, by contrast with the growth of computing resources driven by Moore's Law, roughly a doubling every two years.
We argue that the input of AI researchers is also crucial and its contribution can be objectively estimated.
Consequently, we give a simple argument that explains the 5:1 relation between these two rates.
We then discuss the application of this argument to  
different output measures and compare our analyses with predictions based on machine learning scaling laws proposed in existing literature.
Our quantitative framework facilitates understanding, predicting, and modulating the development of these important technologies.
\end{abstract}

\newpage
\pagenumbering{arabic}

\section{Introduction}

The rapid advance of artificial intelligence 
(AI) and machine learning (ML) is taking even the experts by surprise. 
Given its massive costs and potential impact on so many human activities, it is important to understand the factors which control this progress. 
Much discussion of this topic asserts that computational resources are the dominant factor, and their exponential growth (Moore's First Law, \cite{Moore1965}) is the primary driver of this progress ({\it e.g.}, \cite{Sutton2019}). Is this the only relevant factor?  
Can one formalize this relation and make it quantitative?
What can we say about the future? 

We begin by bringing together a variety of input measures, taking into account computational hardware resources as well as human intellectual work. 
We then standardize and compare a variety of output measures, starting from the traditional publications and patents.  Additionally, popular ML benchmarks provide objective measures for specific ML models,
but do not individually capture the overall growth of the field.  To address this we construct a new, exhaustive index, which we call the Aggregate State of the Art in ML (ASOTA) Index, and validate it by comparison with the other output measures.
The ASOTA Index is defined in terms of ML benchmarks in a way which respects their basic
properties, but unlike individual benchmarks can be continued indefinitely into the future to include yet undiscovered 
advances.

To facilitate understanding, prediction and rational resource allocation, we then
develop and contrast two models of the relationship between inputs and outputs: an approach centered on the concept of a production function such as \cite{CobbDouglas1928} that may be understood as a mechanism which combines inputs like computational resources or AI developers' time and produces outputs such as new ML models, and
a framework based on the ML scaling laws
developed in \cite{CortesEtAl1993,HestnessEtAl2017,KaplanEtAl2020} and many other works that describes 
an empirical
``black-box'' relationship between compute and ML model performance measures.
We quantitatively confirm the belief that Moore's Law is the dominant factor driving progress in AI, 
but that its role is much more nuanced than traditionally assumed
and is better captured by the former model. In particular, this highlights the contribution of human intelligence to pushing the AI frontier further.

For previous work on the productivity and costs of computing, see \cite{Nordhaus2007,RuppSelberherr2011}; 
a related discussion of the dynamics and drivers of technological progress can be found in \cite{BloomEtAl2020} and \cite{FarmerLafond2016}.\footnote{The economics
literature has been long interested in the role of AI/ML development (and automation more generally): in its use as a method for discovering other methods \cite{CockburnEtAl2019,AgrawalEtAl2019}, in its potential for self-improvement (following \cite{vonNeumann1966} and \cite{Good1966}, \cite{BrynjolfssonMitchell2017,BrynjolfssonEtAl2019}), and most widely in its impact on economic growth, structural change and  inequality, with a particular concern if/when AI is a substitute or complement to labor, in other words about automation vs. augmentation (see \cite{BrynjolfssonMcAfee2014,Autor2015,Autor2019,BrynjolfssonMitchell2017,AcemogluRestrepo20018,AcemogluRestrepo2022,Freeman2015,Freeman2018,BrynjolfssonEtAl2019,AghionEtAl2019,Nordhaus2021,AgrawalEtAl2023}), including the themes of Artificial General Intelligence, singularity and existential risk \cite{Bostrom2014,Nordhaus2021,Jones2023,KorinekSuh2024}.}
However, the existing literature lacks a satisfactory measure of the aggregate AI progress, does not provide suitable data on usable computational resources, and does not 
factor in
the role of labor.  All of this is crucial for understanding the determinants behind progress in AI. 
We discuss these gaps in more detail below, and our work aims to address them.

\section{Measures of inputs and outputs}

Moore's First Law is generally stated as ``the number of transistors in an integrated circuit doubles about every two years.''  While this is only approximate, and it leaves out factors such as the speed of computation (which also increased), this rough rate of exponential growth also holds for FLOP/sec per dollar \cite{Wikipedia2024}. 
We define the stock of available computational resources by multiplying the prices FLOP/sec/\$ by the quantity of actual monetary investment in computing in a given time period (one year). Accumulating these investments over time and properly accounting for their depreciation, we obtain a time series $K_t$ of total computational capital plotted in Figure \ref{f:klp}.\footnote{We rely on official US statistics on investments and depreciation. See Supplement for details, including a plot of ingredient data series.} It can not be emphasized enough that the spectacular growth in $K_t$ is almost entirely driven by the exponential decline in the price of FLOP/sec, given much more modest dynamics in investments and the high depreciation rate. 

\begin{figure}
\centering
\includegraphics[width=0.99\textwidth]{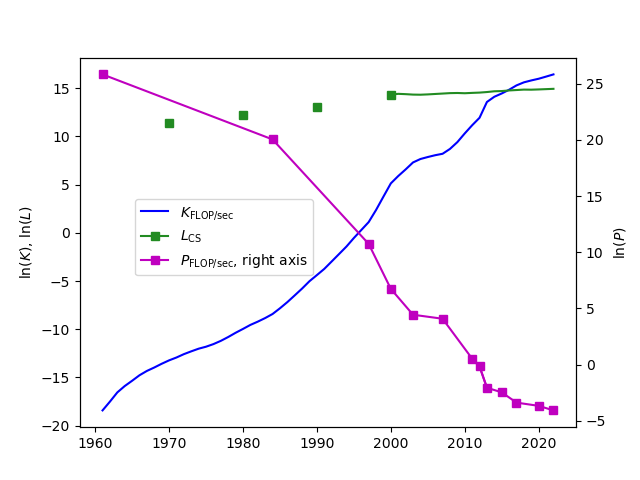}
\caption{Capital (with the price of FLOP/sec) and labor used in the AI/ML technologies sector.\\
{\small[Sectoral boundaries are described in the Supplement. Variables are defined as follows: $K_\textrm{FLOP/sec}$ --- capital stock (in PFLOP/sec, accounting for depreciation); $L_\textrm{CS}$ --- labour in the CS-related occupation (in persons); $P_\textrm{FLOP/sec}$ --- price (US\$ per GFLOP/sec, deflated to 2017 price level).]}}
\label{f:klp}
\end{figure}

To obtain the computational resources devoted to AI development one would need to further multiply this figure by the fraction of total computational resources put into AI research activity, denoted $\phi_\textrm{AI}$.
Here we take the fraction $\phi_\textrm{AI}$ to be constant, an assumption we critically examine later.
\footnote{Note that measuring capital in units directly relevant for production is more defensible than the more usual approach in economics of measuring and aggregating different types of capital in monetary terms.}

\begin{figure}
\centering
\includegraphics[width=0.99\textwidth]{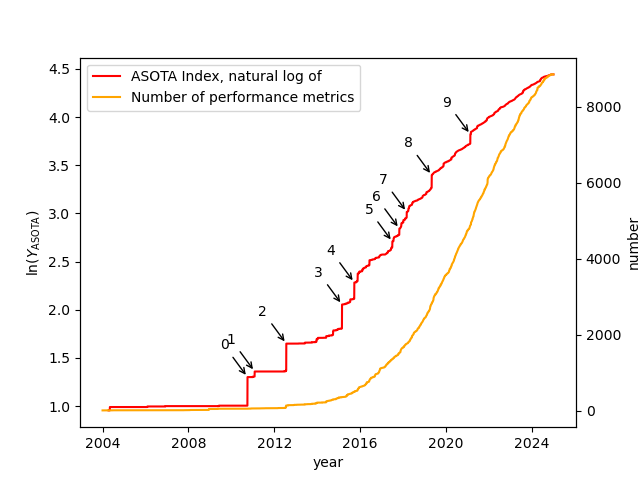}
\caption{Aggregate State of the Art in ML Index.\\
{\small[The number of performance metrics is the number of ML task-dataset combinations available. The Aggregate SOTA Index measures the expansion of the number of ML task-dataset combinations and improvement in their performance metrics. It uses 8858 valid task-dataset combinations available. Computed at the daily frequency, logarithm of the Index reported, 2009 standardized to 1.\\ \footnotesize Annotated increments of the Index (additionally reporting the number of  
combinations with an improvement, and a representative example):
\ \\ (0) 1, including {\tt \lstinline|unsupervised-dependency-parsing-on-penn|};\\
(1) 1, including {\tt \lstinline|video-quality-assessment-on-msu-sr-qa-dataset|};\\
(2) 3, including {\tt \lstinline|atari-games-on-atari-2600-montezumas-revenge|};\\
(3) 15, including {\tt \lstinline|atari-games-on-atari-2600-star-gunner|};\\
(4) 28, including {\tt \lstinline|3d-human-pose-estimation-on-human36m|};\\ 
(5) 10, including {\tt \lstinline|atari-games-on-atari-2600-asteroids|};\\
(6) 9, including {\tt \lstinline|image-generation-on-lsun-bedroom-256-x-256|};\\ 
(7) 30, including {\tt \lstinline|code-generation-on-wikisql|};\\ 
(8) 43, including {\tt \lstinline|machine-translation-on-wmt2016-english-german|};\\ 
(9) 11, including {\tt \lstinline|medical-image-segmentation-on-etis|}. 
]}}
\label{f:index}
\end{figure}

\begin{figure}
\centering
\includegraphics[width=0.99\textwidth]{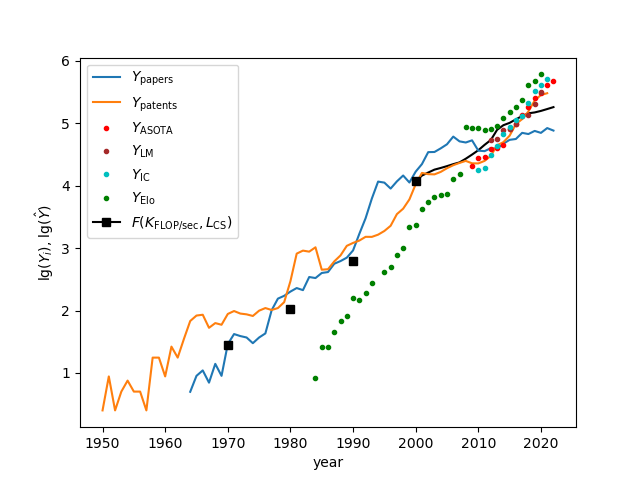}
\caption{Progress in AI/ML technologies.\\
{\small[Theoretical model formulated as $Y_{it} = F_{it} (K_t, L_t) = A_{it} K_t^\alpha L_t^{1-\alpha}$ in logarithms. Output elasticity parameter $\alpha$ calculated from 2017 data. Time-series data is decennial-frequency before 2000, annual after that. Means of output proxy-specific $\ln(A_{it})$ are estimated by OLS, then subtracted from the corresponding proxy series to allow for series' alignment on a common plot. Then, all series standardized to a common metric, chosen to be number of papers published annually, and vertical axis is scaled in terms of decadic (base-10) logarithm of this quantity. Goodness-of-fit measures are $R^2=0.88$ for $Y_\textrm{papers}$ with 26 observations, $R^2=0.93$ for $Y_\textrm{patents}$ with 25 observations, $R^2=0.73$ for $Y_\textrm{ASOTA}$ with 14 observations, $R^2=0.71$ for $Y_\textrm{LM}$ with 9 observations, $R^2=0.66$ for $Y_\textrm{IC}$ with 12 observations, $R^2=0.79$ for $Y_\textrm{Elo}$ with 22 observations.]}}
\label{f:aiml}
\end{figure}

To quantify progress in AI, one can look at traditional measures of research output, such as numbers of published papers and numbers of patents.   One can also look at the performance of state of the art models on standard benchmarks, such as computer chess $Y_\textrm{Elo}$ \cite{Swedish2024}, language modeling $Y_\textrm{LM}$ \cite{TaylorEtAl2003}  
or image classification $Y_\textrm{IC}$ \cite{DengEtAl2009}.\footnote{For a selection of various ``diagnostic'' statistics, see 
\cite{USPTO2020, 
OECD2023}, and lastly \cite{Stanford2023} with their Stanford AI Index.}
However, the list of relevant benchmarks changes with time (e.g., see \cite{OttEtAl2022}). To deal with this we have defined novel ML performance measures that combine many different benchmark performance figures following a systematic procedure, much as is done to construct stock market indices such as the Dow Jones or S\&P 500.  Our chosen measure, called the Aggregate State of the Art in ML Index (in short, Aggregate SOTA or ASOTA Index), is presented in Figure \ref{f:index}.  It captures the improvements in benchmark performance measures, weighing those with a larger number of contributions more highly; and it also captures introduction of new benchmark performance measures (full details on its construction are given in the Supplement). Thus we have various output measures $Y_{it}$ which we seek to relate to $K_t$.  Plotting these measures in Figure \ref{f:aiml}, we see that they are generally consistent and all show a similar exponential growth, which however is far slower than a doubling with every two years.

The idea that growth in compute has a power law relation to progress in AI
has attracted a lot of interest in the AI/ML community, including \cite{OpenAI2018,ThompsonEtAl2020,KaplanEtAl2020}, and \cite{SevillaEtAl2022}. We will discuss some of these publications below, after explaining our own approach to this idea.

\section{A model of research productivity}

We start from the basic economic approach to productivity and growth (e.g., see \cite{Solow1957,LongPlosser1983,Acemoglu2003,BarroSalaIMartin2004}; as well as \cite{Romer1990,AghionHowitt1992}), which focuses on studying exactly this type of relation. In economic terms, computational resources are a form of capital, {\it i.e.}, physical means of production (that is, durable goods which are used for producing other goods). If information and information services are goods, then machines which process information surely qualify.  However, up to the present day, capital does not produce anything by itself: it must be employed by labor.  Indeed, AI researchers are an essential part of the discussion, contributing their effort and cognitive, intellectual resources to the activity.

The simplest quantity expressing this input factor is the number of people who contribute to AI research.
Thus, we measure $L_t$ as the number of people employed in occupations relevant to AI research (see Supplement for details). As motivation, one can argue in the spirit of \cite{SinatraEtAl2016} that by and large, the rate of scientific idea creation can be understood mechanistically as the number of scientists multiplied by a constant discovery rate. 
While this abstracts away details of organization and heterogeneity of the labor force, 
this is justified along the lines  
of \cite{Cohen1991}. 
In modern economics, labor is often viewed as a form of human capital, which captures the fact that different persons can have very different levels of productivity.\footnote{In contrast to physical capital, human capital includes employee knowledge, skills, education (and good health). For a recent review, see \cite{Deming2022}. For a macroeconomic perspective, see the classical \cite{Lucas1988}. In connection to automation, see \cite{Autor2015,AcemogluRestrepo20018,AcemogluRestrepo2022}.} 
Much like capital $K$ whose supply depends on the dynamics of investments and depreciation, human capital $L$ depends on demographic factors such as age distribution and educational attainment.
However, as is common in the literature, we do not account for these factors explicitly, focusing only on the overall headcount (i.e., births/deaths as well as labor market participation decision) and assuming the rest is captured by a fixed multiplicative factor that can be absorbed into other terms of the equation. The plausibility of this assumption is supported by the fact that real wages in the occupations we are focusing on 
maintained a constant premium over the aggregate wage level since 1970 (as shown in the Supplement).

These factors $K$ and $L$ combine to produce new ideas and knowledge.  
In economic terms and for our purposes, new knowledge is an ``output'', denoted $Y_t$, of some process specified by a production function, $Y = F(K, L)$.\footnote{Ultimately, $Y_{it}$ for $i \in \{ \textrm{papers}, \textrm{patents}, \textrm{ASOTA}, \textrm{LM}, \textrm{IC}, \textrm{Elo} \}$ is itself an input for producing, say, medical diagnostics in the health industry or language translation services in the media industry further down the supply chain. Formally, in such applications our production function's output $Y_{it}$ is downstream production function's input, becoming a component of aggregate capital $K_{\textrm{agg},t}$ or technical productivity $A_{\textrm{agg},t}$ that are utilized to produce output $Y_{\textrm{agg},t}$.}  A popular specification for a production function is due to Cobb and Douglas \cite{CobbDouglas1928}:
\begin{equation}\label{e:cobbdouglas}
Y_t = A_t K_t^\alpha L_t^{1-\alpha} .
\end{equation}
The parameter $\alpha$ is the output elasticity with respect to $K_t$ (and analogously $1-\alpha$ for $L_t$), while the variable $A_t$ is an unobserved measure of productivity (which may also capture factors beyond capital and labor, and additionally takes care of the scale and terms of measurement)\footnote{In economic literature, the total factor productivity or technical change term $A_t$ is usually taken as either (i) a residual, (ii) as a stochastic process with an estimated or postulated trend and an added shock, or (iii) endogenously determined variable with explicit dependence on  
inputs such as researchers or lab equipment.}.  This form can be motivated in several ways: as an emergent outcome from plausible assumptions about the underlying microeconomic processes, see \cite{Houthakker1955} and \cite{Jones2005}; or merely as an atheoretic aggregator function exhibiting reasonable economic properties.

In modern economics, every activity is understood as a consequence of the agents' optimization under assumptions about their economic environment. In our case of R\&D production, this amounts to managers using the optimal combination of capital and labor, as well as owners of capital and the labor force being paid competitive rents and wages.  It is a textbook exercise to show that given Eq. (\ref{e:cobbdouglas}), the optimal distribution of the produced revenues is to assign a fraction $\alpha$ to capital and $(1-\alpha)$ to labor.  This allows inferring $\alpha$ indirectly from available data, carefully interpreted.  We thus consider US industry-level statistics, and take the fraction of the corresponding industry's output (more precisely, ``value added'') spent on payroll (``compensation of employees'') as our estimate of $(1-\alpha)$.\footnote{Some of the progress in AI performance is not due to more sophisticated ML models but rather due to algorithmic efficiency of implementing the same models with lower utilization of (primarily) computational resources. This is basically an improvement of the software used, and from economic standpoint software is treated as capital. Since in our case capital is measured as available computational resources in terms of FLOP/sec, algorithmic efficiency is a good example of capital-augmenting technical progress (see more on this below).}

It turns out that whereas for the US economy as a whole one finds $\alpha\sim 0.45$, for research and development organizations one finds $\alpha\sim 0.20$ (see Supplement for calculation details).
Thus Eq. (\ref{e:cobbdouglas}) with $\alpha=0.20$ and residual term $A_t$ is a model for output $Y_t$ over multiple time periods $t$ with a single free parameter.\footnote{
In more detail, we take the logarithm of Eq. (\ref{e:cobbdouglas}) and regard $\log A_t$
as a fixed-variance stochastic process, so we 
estimate its mean value using ordinary least
squares.  This estimation procedure merely ensures the appropriate vertical intercept of the fitted curve.
}
Returning to Figure \ref{f:aiml}, we have also plotted this model's predictions given the constructed historical series of $K_t$ and $L_t$. These predictions are good over a long time scale and are a much better fit than a doubling every two years. It is worth reiterating that we obtained $\alpha$ relying on economic theory and (plausibly uncorrelated) data, instead of optimizing it so as to fit the data $(Y_t,K_t,L_t)$. 
{We notice that papers and patents mostly exceed the model predictions in the earlier period, but usually undershoot the model in the last decade; while the opposite pattern pertains to ML performance measures, chiefly Elo measure.  Interestingly, ML models' performance on standard datasets, when they became available less than two decades ago, exhibit dynamics very similar to that of patents.  Lastly, note that ML models' performance, patents as well as our own $Y_\textrm{ASOTA}$ grow more quickly in recent times, a point we will return to below.}$^,$\footnote{The periods 1974--1980 and 1987--2000 (sometimes split into two periods 1987--1993 and early 2000s, often without the latter subperiod), which are colloquially referred as ``AI winters'' due to intellectual stumbling blocks and reduced funding, do not seem like outliers from our framework's perspective.}

\section{Differences across output measures and over time}

The relation Eq. (\ref{e:cobbdouglas}) with $\alpha=0.20$ holds to a reasonable approximation for all of the output measures and over all time.  However, a closer examination suggests that there may be significant heterogeneity in the data.  While the output measures available for the full timespan
($Y_\textrm{papers}$ and $Y_\textrm{patents}$) double every $10$ years, the measures available for shorter periods grow considerably faster.
The machine learning benchmarks such as $Y_\textrm{ASOTA}$ have doubling times around $2.5$ to $3$ years.
Notably, this also includes computer chess performance $Y_\textrm{Elo}$, for which we have data from well before the modern AI era (although the early systems did not rely on ML).  On the other hand, the more traditional output measure $Y_\textrm{patents}$ also grew equally fast since 2012. 

There are plausible ways to rationalize these 
discrepancies. First, there are clear differences between the output measures.
In contrast to papers and patents, ML benchmarks $Y_\textrm{LM}$ and $Y_\textrm{IC}$ are bounded and cannot grow forever ({\it e.g.}, modern models already demonstrate a perfect score on the famous MNIST task \cite{LecunEtAl1998}, see \cite{OttEtAl2022} for a general perspective on this).  Our $Y_\textrm{ASOTA}$ is also based on bounded measures but it is constantly expanding with new tasks and datasets, while $Y_\textrm{Elo}$ is not an absolute performance measure as the ones above but measures relative performance of ML models. Also, advances in algorithmic efficiency that are unrelated to AI are possibly more relevant for improving the performance on ML benchmarks than for publishing academic papers or obtaining patents.

Second, there are good reasons to believe that the computational resources $K_t$ used for AI research increased much more quickly in recent years than implied by our estimate based on investments made for the entire computing sector and the assumption of constant fraction $\phi_\textrm{AI}$.  For instance the revenue of NVIDIA, the leading producer of Graphical Processing Units (a reasonable proxy for computational resources used in modern AI)  constituted 0\% of total US investments in computer equipment at the firm's outset in 1996, 1\% in 2001, 3\% in 2010, and 18\% in 2022.  This suggests significant growth of $\phi_\textrm{AI}$ in recent years.

More evidence for this comes from the recent works \cite{OpenAI2018,SevillaEtAl2022}.
These works use a different measure of computational resources, the compute $C_t$
used to train an individual ``milestone'' ML model which advanced the state of the art at time $t$.  
While different from our measure $K_t$ of total computational resources, the two are plausibly related (beyond just the obvious $K_t \ge C_t$). 
These works find more structure in their time series than we find in our $K_t$.  In particular, they find different rates of growth of compute over time, with faster rates in later time periods.
If we were to grant the same growth rate for $K_t$ between 2010 and 2022 as found for $C_t$ in \cite{SevillaEtAl2022},
the relation Eq. (\ref{e:cobbdouglas}) with $\alpha=0.20$
would better fit $Y_\textrm{ASOTA}$ and the other measures (though not $Y_\textrm{papers}$).\footnote{
  The relevant passage in \cite{SevillaEtAl2022}: ``We identify an 18-month doubling time between 1952 and 2010, a 6-month doubling time between 2010 and 2022, and a new trend of large-scale models between late 2015 and 2022, which started 2 to 3 orders of magnitude over the previous trend and displays a 10-month doubling time.''  
}

Finally, several decades of autocorrelated annual observations is, statistically speaking, a small sample.  This paucity of data
motivates the use of a minimal model saving on free parameters.  As we explained, the single parameter $A$ of our model is required to
take into account the scale and terms of measurement, while $\alpha$ was determined by other economic considerations.
This extremely parsimonious model can be viewed as a ``macro'' summarization of the diverse output measures.

\section{Machine learning scaling laws}

Let us turn to an ML-based approach to our questions.  The computer science community is interested in understanding the relation between resources and performance of ML models, which turns out to follow \textit{scaling laws}.  Generally, this relation is studied for particular tasks, for which the usage of compute can be precisely defined
and varied in controlled experiments.

In modern AI, the operation with the largest compute requirement is the training of a model.  
Consider a large language model (LLM).
Its basic task is to continue a text; in other words given a sequence of words, it must predict the word most likely to follow the given sequence.  
An LLM is trained to do this by going through an entire
corpus of text and for each word, slightly varying the LLM parameters to increase the probability of correctly predicting it.  This suggests, and it is indeed the case, that the compute required to train a standard ML model is $C \sim D \cdot P \cdot T$, where $C$ is measured in FLOPs, $D$ is the size of the dataset, $P$ is the size of the model (usually the number of parameters), and $T$ is an order-one factor counting the number of passes over the training data and other particulars.\footnote{To relate this to our previous definitions, the compute $C_j$ devoted to model $j$ in a given year is some share $\varphi_j$ of available computational resources $S \cdot K_t$ with $S$ denoting number of seconds in a year. Note that the definition of $C$ by construction incorporates the quantity of data $D$, which is recognized as a very important factor in AI progress, see \cite{LeCunEtAl2015,HartmannHenkel2020}.} 
This relationship holds for a very wide range of models and tasks, and since 
the right-hand-side terms in the relationship are independent of $C$ and $Y$, they can be treated as parameters exogenously controlled in experiments (within the limits of available resources).

Scaling laws of the form $Y = C^{\alpha'}$ have been proposed as general properties of ML systems \cite{CortesEtAl1993,HestnessEtAl2017,KaplanEtAl2020}; they are supported by both the empirical evidence (via training a series of ML models of the same form using different compute $C$ and measuring the performance $Y$, {\it e.g.} \cite{KaplanEtAl2020}) and theoretical arguments (see \cite{BahriEtAl2021,MaloneyElAl2022}).\footnote{There are similar laws for the joint dependence on dataset size $D$ and model size $P$.  A typical form \cite{HoffmannEtAl2022} relates $L$ (a loss
function such as error rate) to $D$ and $P$ as $L = L_{min} + B/D^\beta + G/P^\gamma$, where $L_{min}$ is the minimal possible loss for the task.  One can use such a law to
optimize the division of resources between dataset and model size.
Interestingly (and encouragingly), doing this recovers a $Y = C^{\alpha'}$ scaling law:  generally $\gamma \sim \beta$, and taking $C \sim D \cdot P$, at the optimal allocation for $D$ and $P$ at fixed $C$, we obtain $D \propto P$ and $\alpha' = \beta / 2 = \gamma / 2$.}. However, while the proposed form of this scaling law is well accepted in the literature, the exponent $\alpha'$ is not universal and varies depending on the task, model and even dataset. For a wide selection of models, its value ranges as 
$0.05\lesssim\alpha' \lesssim 0.15$, as found by \cite{KaplanEtAl2020} and \cite{HoffmannEtAl2022} as well as \cite{ThompsonEtAl2020}, with the latter pointing out that the computing power required by this law with $\alpha'\ll 1$ could become a major obstacle to progress in AI.\footnote{We should mention that some works,
for example \cite{NeumannGros2023} focusing on reinforcement learning models, find $\alpha'$ as large as $0.6$.}

\section{Comparison of the two frameworks}

The parameters $\alpha\sim 0.2$ of the economic model and $\alpha' \sim 0.1\pm 0.05$ of the
ML scaling laws both govern the relation between input compute and output performance, 
and have similar numerical values, suggesting that they can be directly compared.
But before we do this, let us explain the differences between the frameworks.

First, the two frameworks describe different relations between inputs and outputs. 
Our economic framework is concerned with the overall development of AI knowledge, including new techniques and ever improving models, as a function of the total computational resources and labor employed. By contrast, an ML scaling law pertains to a specific model applied to a specific task. Since the former involves many instances of the latter, selected and combined by AI researchers, and different models are not truly exchangeable, it is not clear whether the two frameworks should give the same value of the exponent.  

Second, the measurement approach in the economics framework ascribes all the output share beyond labor to capital, while it can be argued that there are additional productive factors relevant for AI progress, such as energy.\footnote{See, e.g., \cite{DeVries2023}.  This is also implied in a framework like KLEMS due to Jorgenson \cite{JorgensonEtAl1987}).}  
Thus, our estimate $0.2$ could be viewed as an upper bound on $\alpha$. 
Third, in the economics framework the method of calculating this parameter deals with a significantly broader definition of industry related to AI technologies than is the case for specific ML applications considered above. Fourth, there is meaningful variation over time of the inferred $\alpha$ in the economic applications (see studies cited 
in the Supplement) and in the ML framework across different studies/applications (cited above).  Given all this, it is noteworthy that the two approaches produce remarkably similar measurements, $\alpha\sim 0.2$ versus $\alpha'\sim 0.1$ ({\it cf.} the economy-wide value of 0.45).

\section{Conclusions}

Our main results are twofold.  First, we provide a dataset of measures of AI research output, including a new
Aggregate State of the Art in ML (or ASOTA) index, which can be continued into the future and provide a solid
foundation for research in this area.  Second, we show that the Cobb-Douglas production function with the
standard ``capital'' input factor replaced by a computational resources factor, with an output elasticity of 
$\alpha=0.2$, fits these output measures to reasonable accuracy
over the span of five decades.
The extreme simplicity of this ``minimal economic model'' and
the absence of free parameters (recall that $\alpha=0.2$ was obtained from
another, entirely different data source) is remarkable. 

The similarity of the ``macro'' minimal economic model to the ```micro'' machine learning scaling laws,
both in form and in the values of the scaling exponents $\alpha\sim\alpha'$,
is a further evidence for its validity.  This also holds out hope for the development
of structural models which relate the two 
levels of explanation.

Our model can be seen as a 
quantification of the idea that Moore's Law has been the primary driver
of progress in AI (and computer science more generally) \cite{Sutton2019}.
Many authors have argued that Moore's Law has slowed down in recent years (\cite{RuppSelberherr2011,Williams2017,Flamm2017}, but also see \cite{Nordhaus2007,Gargini2017}),
and this is consistent with the data on FLOP/sec prices and our estimate of computational capital in Figure \ref{f:klp}.
To complement these calculations,
an important open question is to better estimate the compute resources actually devoted to AI research
(or equivalently the fraction $\phi_\textrm{AI}$).  There are reasons to think that this grew substantially
over the period 2012 to the present, perhaps explaining the differences between output measures visible in
Figure \ref{f:aiml}.  Since $\phi_\textrm{AI}\le 1$, such growth cannot compensate for a slowdown in Moore's Law forever.

Compared to previous work, we feel the most 
underappreciated point highlighted by the minimal
economic model is the importance of the labor input factor.  Its high elasticity $1-\alpha=0.8$ 
means that increases in labor (or human capital) translate almost fully
into research output.
This signals a larger need for highly skilled researchers, especially as Moore's Law slows down.
This will help address the problem of jobs lost to AI and create new well paid positions.

One can ask whether there are other important input factors.  For example, it might be that $A_t$
in Eq. (\ref{e:cobbdouglas}), the ``research productivity,'' is growing exponentially.  
Moreover, it is widely expected that the application of AI will drive improvements in productivity in many industries. Taking this logic further, we need to also understand the effect of AI progress on AI research productivity itself.\vskip6pt

\bibliographystyle{plain}
\bibliography{ProgInAI+.bib}

\appendix

\renewcommand{\thefigure}{S\arabic{figure}}

\clearpage\newpage
\begin{center}
    \LARGE \textbf{Supplementary Material}
\end{center}
\bigskip
\begin{center}
    \large \textbf{Additional Details on Methods}
\end{center}

\bigskip\bigskip
\begin{center}
    \textbf{Production function, production factors and output}
\end{center}
Let us repeat Eq. (1)  
from the main text:
\[ Y_t = A_t K_t^\alpha L_t^{1-\alpha}.  \]
Our model posits that the stock of the two inputs $K$ and $L$ are combined to produce the flow of output $Y$ (similarly to how economists model the production of goods or the provision of services given the corresponding factors of production).

Next, we discuss the sources of data and methods to estimate quantities of interest necessary for utilizing the Cobb-Douglas production function. Output elasticity parameters $\alpha$ and $(1-\alpha)$ (assuming Constant Returns to Scale, which is a standard approach in economic literature) are sourced from U.S. BEA statistics (more on which below); the available amounts of computational and cognitive/intellectual resources are also taken form U.S. official statistics (see below); while the measures of performance improvements come from various publicly available datasets (see below).

\begin{figure}
\centering
\includegraphics[width=0.99\textwidth]{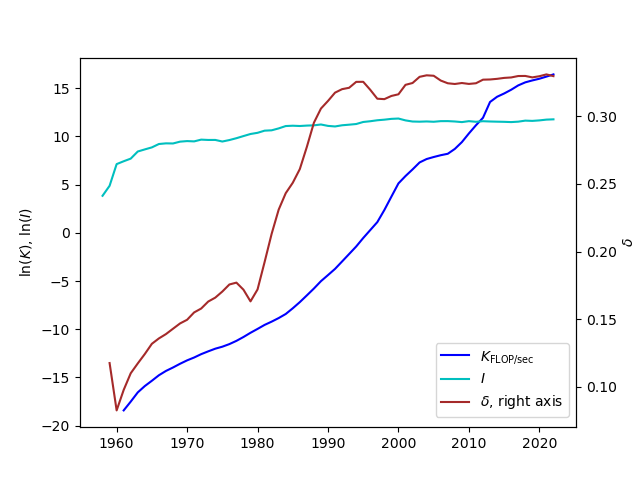}
\caption{Investments, capital and its depreciation for the AI/ML technologies sector.\\
{\small[
Variables are defined as follows: $K_\textrm{FLOP/sec}$ --- capital stock (in PFLOP/sec, accounting for depreciation); $I$ --- investments in the corresponding industry (in US\$mn, deflated to 2017 price level);
$\delta$ --- depreciation rate in the corresponding industry (in terms of share).
]}}
\label{f:kidelta}
\end{figure}

Capital stock is calculated as investment flows in terms of FLOP/sec accumulated over time and accounting for depreciation (broadly following the so-called ``perpetual-inventory method''). Formally,
$$K_t := (1-\delta_t) K_{t-1} + (1-0.5\delta_t) I_t,$$
where $I_t$ is investments in terms of FLOP/sec made available at year $t$, $\delta_t$ is the depreciation rate at $t$, and $K_t$ is the total amount of capital available in the economy in terms of FLOP/sec. $I_t$ is measured as Investment in Private Fixed Assets for Computers and peripheral equipment sourced from BEA divided by Computing hardware costs in USD per FLOP/sec from Wikipedia (after log-linear interpolation), with prices deflated appropriately by GDP Deflator from the Federal Reserve Bank of St. Louis database. 
Depreciation rate is the implicit depreciation rate calculated using the formula given above with the same $I_t$ as there but with $K_t$ being Historical-Cost Net Stock of Private Fixed Assets for Computers and peripheral equipment from BEA.
See Figure \ref{f:kidelta} for the plot of the constructed series.

Also, one might argue that this should be multiplied by the fraction of computational resources devoted to AI research, but it is not clear this is well-defined as other computing research can also benefit AI.

Labor input is taken as the Number of persons employed as Computer systems analysts and computer scientists from IPUMS.

Note that in the case of both capital an labor, we do not assume that the above statistical series measure the relevant factors exactly (in fact, they are overestimating them since the categories used are broader than AI and ML); but, due to the presence of $A_t$ and constant $\alpha$, our approach remains valid as long as the measures we use differ from our required measures by a factors of proportionality (possibly different for capital and for labor).

As performance measures, we take the annual Number of papers published on ``Artificial Intelligence'', the annual Number of patents published on ``Artificial Intelligence'', as well as several accepted ML benchmarks: a standard Language Modelling task on a popular dataset called Penn Treebank (originally measured in terms of ``Perplexity'' and mapped to $[0,1]$, with higher value signifying better performance, taken as the level achieved at a given period), a classical Image Classification task on a standard dataset ImageNet (originally measured in terms of ``Top 5 Accuracy'' and mapped to $[0,1]$, with higher value signifying better performance, taken as the level achieved at a given period), a popular Chess Engine performance measure (originally measured in terms of Elo rating and divided by the calibration constant 400, taken as the level achieved at a given period), and our own broad measure called Aggregate SOTA Index (described below, taken as the level achieved at a given period). To express all of these measures in logarithmic terms, a natural logarithm is applied to papers, patents, Language Modelling and Image Classification measures (Elo rating and Aggregate SOTA Index are initially defined already in logarithmic terms).\footnote{Expressing $Y_{it}$ in levels may pose difficulties for production function's interpretation. Specifically, when $K_t < K_{t-1}$ and $L_t < L_{t-1}$, in order to ensure a non-decreasing stock of developed AI technologies (as represented by, e.g., the level of Aggregate SOTA Index) we would be forced to conclude that unobserved productivity rises, $A_{it} > A_{i,t-1}$. 
}

Output elasticity parameters of capital and labor $\alpha$ and $(1-\alpha)$, respectively (equivalently in our specification, the capital and labor shares of income), are calculated using official U.S. industry-level statistics. We take the ``Scientific research and development services'' industry's data on Compensation of employees as well as on Value added from Input-Output Accounts Data that U.S. Bureau of Economic Analysis regularly provides\footnote{These data are the building blocks of the official estimates of gross domestic product and are also used for estimating the effects of various policies and regulations, such as tax proposals.} and, given the production function specification as well as assuming competitive capital and labor markets (which is a common assumption in such cases), we can obtain a measure of labor share of income $(1-\alpha)$ by taking the ratio of Compensation of employees over Value added.\footnote{Since Marschak and Andrews \cite{MarschakAndrews1944} it is known that a na\"{i}ve approach of estimating these parameters empirically by running an OLS regression of output variable on factor input variables is inconsistent due to endogeneity/simultaneity problem: factor inputs in general are not independent from unobserved productivity, e.g., firm managers may respond to new information about productivity and technical progress, and adjust factor usages accordingly.} Note that for calculating factor shares we are relying on the latest industry data released by the BEA. The reason is that, in light of industry-level evolution over the history and the resulting changes in the definitions of industry composition, it is difficult to construct a consistent measure over our whole sample period.\footnote{This introduces the obvious risks of mismeasurement; for instance, it has been observed that the \textit{aggregate} capital share $\alpha$ was lower several decades ago and has been increasing since then (e.g., see \cite{ElsbyEtAl2013}; \cite{KarabarbounisNeiman2013}; \cite{Rognlie2015}). So, if the research industry we are interested in exhibited a similar trend, by using relatively recent data throughout our study we may be overestimating the contribution of computational resources and---given their fast expansion---the pace of AI/ML progress implied by our model for the earlier period (in other words, the model-implied slope could be a little too steep in the earlier periods of the sample).}$^,$\footnote{One may be concerned that violation of perfect competition on the market for computational resources due to prevalence of a few large producers leads to high price markups and subsequent overestimation of the value of capital share $\alpha$ throughout the whole sample. However, in this study we do not estimate $\alpha$ directly from capital factor payments, calculating it instead as a remainder after subtracting the labor share --- i.e., using the $(1-\alpha)$ relation given by the CRS property.}

Taking the logarithms of both sides of the equation, we estimate the mean of $A_t$ by minimizing squared residuals between two sides of the production function equation (which automatically also provides us with a measure of $A_t$ dynamics as the residuals, i.e., the difference between the two sides).
Note that with this approach we estimate just one parameter per output proxy (basically, we have a single regression line with different intercepts).

\bigskip\bigskip
\begin{center}
    \textbf{ML performance Indices}
\end{center}
We use different models' performance results on all ML benchmark tasks and all datasets available from Papers With Code database. It starts in 1998, with first metrics improvements in 2004, and the number of available task-dataset combinations reaching 50 in 2009; currently it contains 
8858 valid task-dataset combinations (with 1106 of them having at least 10 model performance entries).

Then we construct several daily-frequency aggregate meta-measures. Below, for a performance measure $X_{it}$ for task-dataset combination $i$ at date $t$, a rate of improvement $Z_{it}$ is calculated as $Z_{it} := (\sfrac{X_{it}}{X_{i,t-1}^*}) - 1$ when the metrics is of the accuracy type, and $Z_{it} := 1 - (\sfrac{X_{it}}{X_{i,t-1}^*})$ when the metrics is of the loss type, where $X_{it}^* := \max_{\tau \in \{1, \ldots, t\}} X_{i\tau}$. Specifically, we calculate:
\begin{enumerate}
    \item Number of Metrics included as of period $t$: it is defined as $N_t := \sum_i \mathds{1}(X_{it}^*>0)$, i.e., task-dataset combinations with at least 1 entry by a given date;
    \item Equal-Weighted Index: define $\Delta^\text{EW}_t := \sfrac{\sum_i Z_{it}}{\sum_i \mathds{1}(Z_{it} > 0)}$, average rate of improvement on a given date that is weighted equally, and the index is a cumulative product of $(1+\Delta^\text{EW}_t)$ over $t$;
    \item Activity-Weighted Index: define $\Delta^\text{AW}_t := \sfrac{\sum_i Z_{it} \sum_{\tau=1}^t \mathds{1}(X_{i\tau}>0)}{\sum_i \mathds{1}(Z_{it} > 0) \sum_{\tau=1}^t \mathds{1}(X_{i\tau}>0)}$, average rate of improvement on a given date that is weighted by the number of performance entries so far, and the index is a cumulative product of $(1+\Delta^\text{AW}_t)$ over $t$;
    \item Equal-Weighted Expanding Index: define \newline$\Delta^\text{EWE}_t := \left(\sfrac{\sum_i Z_{it}}{\sum_i \mathds{1}(Z_{it}>0)}\right) \times \left(\sfrac{\sum_i \mathds{1}(Z_{it}>0)}{\sum_i \mathds{1}(X_{it}^*>0)}\right)$, average rate of improvement on a given date that is weighted equally multiplied by the number of improvements over the number of metrics included so far, and the index is a cumulative product of $(1+\Delta^\text{EWE}_t)$ over $t$;
    \item Activity-Weighted Expanding Index: define \newline$\Delta^\text{AWE}_t := \left(\sfrac{\sum_i Z_{it} \sum_{\tau=1}^t \mathds{1}(X_{i\tau}>0)}{\sum_i \mathds{1}(Z_{it} > 0) \sum_{\tau=1}^t \mathds{1}(X_{i\tau}>0)}\right) \times \left(\sfrac{\sum_i \mathds{1}(Z_{it}>0)}{\sum_i \mathds{1}(X_{it}^*>0)}\right)$, average rate of improvement on a given date that is weighted by the number of performance entries so far multiplied by the number of improvements over the number of metrics included so far, and the index is a cumulative product of $(1+\Delta^\text{AWE}_t)$ over $t$;
    \item Equal-Weighted Renewing Index: define \newline$\Delta^\text{EWR}_t := \left(\sfrac{\sum_i Z_{it}}{\sum_i \mathds{1}(Z_{it}>0)}\right) \times \left(\sfrac{\sum_i \mathds{1}(Z_{it}>0)}{\sum_i \left(\mathds{1}(X_{it}^*>0)-\mathds{1}(X_{i,t-365}^*>0)\right)}\right)$, average rate of improvement on a given date that is weighted equally multiplied by the number of improvements over the number of metrics included during the last year, and the index is a cumulative product of $(1+\Delta^\text{EWR}_t)$ over $t$;
    \item Activity-Weighted Renewing Index: define \newline{\footnotesize $\Delta^\text{AWR}_t := \left(\sfrac{\sum_i Z_{it} \sum_{\tau=1}^t \mathds{1}(X_{i\tau}>0)}{\sum_i \mathds{1}(Z_{it} > 0) \sum_{\tau=1}^t \mathds{1}(X_{i\tau}>0)}\right) \times \left(\sfrac{\sum_i \mathds{1}(Z_{it}>0)}{\sum_i \left(\mathds{1}(X_{it}^*>0)-\mathds{1}(X_{i,t-365}^*>0)\right)}\right)$}, average rate of improvement on a given date that is weighted by the number of performance entries so far multiplied by the number of improvements over the number of metrics included during the last year, and the index is a cumulative product of $(1+\Delta^\text{AWR}_t)$ over $t$.
\end{enumerate}

The motivations for these measures are the following ones. The first measure, Number of Metrics, merely tracks the number of task-dataset combinations available at a given period. The following two measures quantify the rate of performance improvement in available task-dataset combinations relatively to previously achieved metrics levels, weighting them either equally in the case of Equal-Weighted Index or proportionally to the number of performance entries in a given combination for Activity-Weighted Index. The next two measures, Equal-Weighted and Activity-Weighted Expanding Indices, in addition to the rate of improvement in available task-dataset combinations also quantify the rate of growth of new or update of existing combinations relatively to their cumulative total number (the underlying geometric logic implies the comparison of the area of the rectangle formed by average magnitude of improvement in task-dataset combinations and the number of such new or updated combinations \textit{vs.} the area of the rectangle formed by the previously achieved average metrics levels and the total number of previously available combinations). The last two measures, Equal-Weighted and Activity-Weighted Renewing Indices, are similar to two previous ones, but the rate of growth or update of task-dataset combinations considered is calculated relatively to their cumulative number over the last year (preventing obsolete metrics from affecting the results infinitely far in the future).

Our preferred measure is Activity-Weighted Renewing Index. We use a logarithm of it, and standardize the resulting series so as it equals 1 in 2009, when the number of reported task-dataset combinations has reached 50. The resulting series is called in the text Aggregate State of the Art in ML Index.

\bigskip\bigskip
\begin{center}
    \textbf{Data sources}
\end{center}
Below is a list of exact sources of the data used in the study, as well as references to papers that directly provide the data and/or describe the details of these data.

\medskip
\textbf{1. Performance measures.}

Chess Engine performance measure, Elo rating list (year-end leaders). Swedish Chess Computer Association. \newline\url{https://en.wikipedia.org/wiki/Swedish_Chess_Computer_Association#Rating_list_year-end_leaders}.

Language Modelling on Penn Treebank (Word Level), 2014--2021. \newline{\small\url{https://paperswithcode.com/sota/language-modelling-on-penn-treebank-word}}.

Image Classification on ImageNet, 2015--2023. \url{https://paperswithcode.com/sota/image-classification-on-imagenet}.

ImageNet Large Scale Visual Recognition Challenge, 2010--2017.
\url{https://image-net.org/challenges/LSVRC}.

Number of results of types Article, Proceeding Paper, Book Chapters, Book from Web of Science Core Collection for Web Of Science Category ``artificial intelligence'', \url{https://www.webofscience.com}.

Number of patents, annual; Top 10 classifications for query ``artificial intelligence'', Earliest publication date (family), Espacenet, \url{https://worldwide.espacenet.com/patent}.

State-of-the-Art, Papers With Code. \url{https://paperswithcode.com/sota}, (retrieved on 23/11/2024).

\medskip
\textbf{2. Capital.}

Computing hardware costs. Approximate USD per GFLOP/s (2022 prices), available at \url{https://en.wikipedia.org/wiki/FLOPS#Cost_of_computing}.

Computers and peripheral equipment (k3ntotl1ep11); annual, 1925--2022. Table 2.3. Historical-Cost Net Stock of Private Fixed Assets, Equipment, Structures, and Intellectual Property Products by Type. Bureau of Economic Analysis.

Computers and peripheral equipment (i3ntotl1ep11); annual, 1901--2021. Table 2.7. Investment in Private Fixed Assets, Equipment, Structures, and Intellectual Property Products by Type. Bureau of Economic Analysis.

\medskip
\textbf{3. Labor.}

Number of persons employed in 1950--2022; from IPUMS, Occupation (1990 basis), Person weight, Census years.

Number of persons employed as ``64 Computer systems analysts and computer scientists'' in 1970--2022; from IPUMS, Occupation (1990 basis), Person weight, Census years.

Wage and salary income for all employed persons in 1950--2022; means and 99th percentile; from IPUMS, Occupation (1990 basis), Person weight, Census years.

Wage and salary income for persons employed as ``64 Computer systems analysts and computer scientists'' in 1970--2022; means and 99th percentile; from IPUMS, Occupation (1990 basis), Person weight, Census years.

\medskip
\textbf{4. Capital and labor shares of income.}

Compensation of employees. (Aggregate, Scientific research and development services industry.) Industry Economic Account statistics, Use Tables (Use of commodities by industry), 402 Industries, 2017. Bureau of Economic Analysis.

Value added (producer value). (Aggregate, Scientific research and development services industry.) Industry Economic Account statistics, Use Tables (Use of commodities by industry), 402 Industries, 2017. Bureau of Economic Analysis.

\medskip\textbf{5. Inflation rates and interest rate.}

Consumer Price Index for All Urban Consumers: All Items in U.S. City Average. Index 1982-1984=100, Seasonally Adjusted. Monthly Frequency. Federal Reserve Bank of St. Louis.

Gross domestic product (implicit price deflator). Index 2017=100, Not Seasonally Adjusted. Annual Frequency. Federal Reserve Bank of St. Louis.

\bigskip
\textbf{Data source literature references}

\begin{itemize}
\item Kim, Yoon, Yacine Jernite, David Sontag, Alexander Rush. (2016). ``Character-Aware Neural Language Models''. \textit{Proceedings of the AAAI Conference on Artificial Intelligence}, 30 (1).

\item Ruggles, Steven, Sarah Flood, Matthew Sobek, Daniel Backman, Annie Chen, Grace Cooper, Stephanie Richards, Renae Rogers, and Megan Schouweiler. IPUMS USA: Version 14.0 [dataset]. Minneapolis, MN: IPUMS, 2023.

\item Russakovsky, Olga, Jia Deng, Hao Su, Jonathan Krause, Sanjeev Satheesh, Sean Ma, Zhiheng Huang, Andrej Karpathy, Aditya Khosla, Michael Bernstein, Alexander C. Berg and Li Fei-Fei. (2015). ``ImageNet Large Scale Visual Recognition Challenge.'' \textit{International Journal of Computer Vision}, 115: 211–-252.
\end{itemize}

\clearpage\newpage
\begin{center}
    \large \textbf{Additional Text}
\end{center}

\bigskip\bigskip
\begin{center}
    \textbf{Production function}
\end{center}
Cobb-Douglas specification satisfies the mathematical properties of what is known as neoclassical production function: constant returns to scale, positive and diminishing returns to private inputs, Inada conditions on behavior at the extremes. Jones \cite{Jones2005} shows how Cobb-Douglas production function can be derived from microeconomic foundations presuming that techniques for combining capital and labor to produce output are drawn from Pareto distributions (whose shape parameters define the production function's exponent $\alpha$). This production function has some theoretically restrictive assumptions such as constant output elasticity parameters as well as empirical challenges about measurement of its inputs and identification of its parameters; but because of its analytical convenience, theoretically appealing features such as admitting a balanced growth path and a constant positive capital share with technological change not necessarily being labor augmenting as well as satisfactory empirical performance, it is
the workhorse specification for long-term highly-aggregated economic analysis and forecasting (e.g., see \cite{Jones2003} for a concise theoretical discussion and \cite{VanBeveren2012} for an overview of the empirical aspects).

\begin{figure}
\centering
\includegraphics[width=0.99\textwidth]{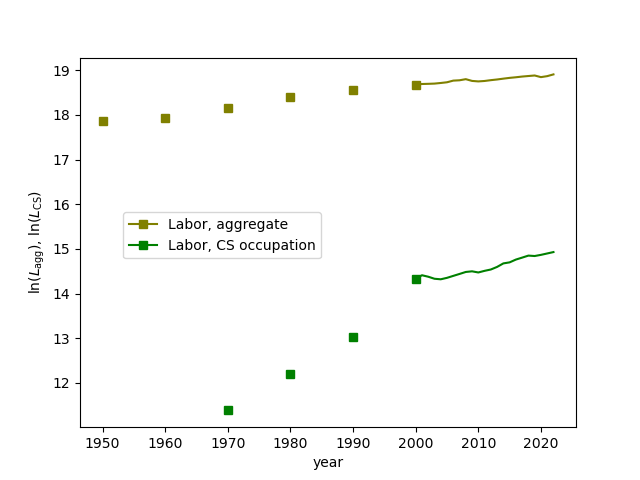}
\caption{Labor used in the AI/ML technologies sector.\\
{\small[Variables are defined as follows: $L_\textrm{agg}$ --- labour in the aggregate economy (in persons); $L_\textrm{CS}$ --- labour in the CS-related occupations (in persons).]}}
\label{f:labor}
\end{figure}

\bigskip\bigskip
\begin{center}
    \textbf{Labor statistics}
\end{center}
From Figure \ref{f:labor}, we can see that the amount of labor in the AI-related sector was growing at a faster rate than in the economy overall, but the growth rate has substantially slowed down since 2000.

\begin{figure}
\centering
\includegraphics[width=0.99\textwidth]{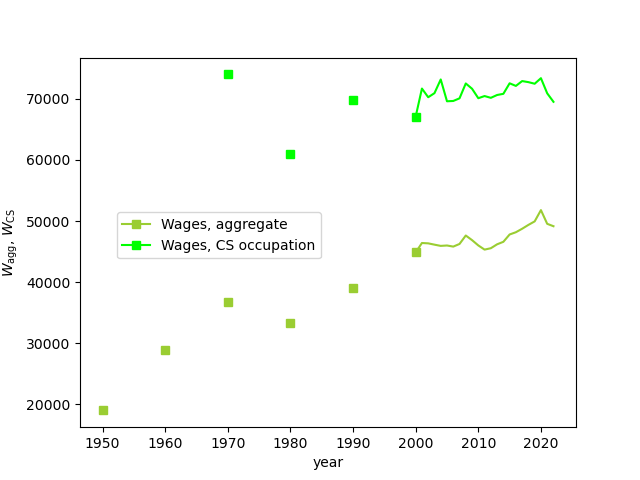}
\caption{Wages paid in the AI/ML technologies sector.\\
{\small[Variables are defined as follows: $W_\textrm{agg}$ --- annual wages in the aggregate economy (average); $L_\textrm{CS}$ --- annual wages in the CS-related occupations (average). Deflated to 2017 price level.]}}
\label{f:wage}
\end{figure}

In our study, all
wages are deflated by the CPI from the Federal Reserve Bank of St. Louis database. In Figure \ref{f:wage} one can see the time series of wages in the AI-related occupations and in the wider economy. Assuming competitive labor markets, the difference in wages is the premium to human capital in the latter sector. Surprisingly, it is remarkably stable. 

\end{document}